\documentclass[12pt,a4paper]{article}
\usepackage{latexsym,amssymb,amsmath,amsthm}

\newtheorem{theorem}{Theorem}
\newtheorem{lemma}{Lemma}
\newtheorem{corol}{Corollary}

\renewcommand{\proof}{\noindent {\it Proof. }}
\newcommand{\eq}{\Longleftrightarrow}

\begin{document}
%ÓÄÊ 519.7, 519.61
\title{Thin circulant matrices and lower bounds on
the complexity of some Boolean operators\footnote{Original text
published in Russian in {\it Diskretnyi Analiz i Issledovanie
Operatsii $($Discrete analysis and operations research$)$}. 2011.
18(5), 38--53.}}
\date{}
\author{M. I. Grinchuk, I. S. Sergeev}

\maketitle

\begin{abstract}
We prove a lower bound
$\Omega\left(\frac{k+l}{k^2l^2}N^{2-\frac{k+l+2}{kl}}\right)$ on
the maximal possible weight of a $(k,l)$-free (that is, free of
all-ones $k\times l$ submatrices) Boolean circulant $N \times N$
matrix. The bound is close to the known bound for the class of all
$(k,l)$-free matrices. As a consequence, we obtain new bounds for
several complexity measures of Boolean sums' systems and a lower
bound $\Omega(N^2\log^{-6} N)$ on the monotone complexity of the
Boolean convolution of order $N$.
\end{abstract}


\section{Introduction}

Hereafter, a Boolean matrix is called {\it$(k,l)$-free} (or thin)
if it does not contain an all-ones $k\times l$ submatrix. In the
case $k=l$ we write simply {\it$k$-free}. Further, assume $2 \le k
\le l$.

An $N \times N$ matrix $(c_{i,j})$ is {\it circulant} (or cyclic),
if either $c_{i,j} = c_{0,(i+j) \bmod N}$ for all $i,j$, or
$c_{i,j} = c_{0,(i-j) \bmod N}$ for all $i,j$.

In~\cite{G} the first author proved the existence of $k$-free
Boolean circulant $N\times N$ matrices of weight\footnote{{\it
Weight} of a (Boolean) matrix is the number of non-zero entries in
it.} $\Omega\left(k^{-4}N^{2-\sqrt{3/k}}\right)$ and obtained
corollaries for the complexity\footnote{The reader can find the
notions of {\it complexity}, {\it depth}, {\it rectifier circuit},
{\it circuit of functional elements} e.g. in~\cite{lu,L}.} of
Boolean sums' systems\footnote{{\it Boolean sum} is a function of
the form $x_1 \vee \ldots \vee x_n$. A system of Boolean sums with
an $N\times N$ matrix $(c_{i,j})$ is a mapping with components
$\bigvee_{j=1}^N c_{i,j}x_j$, $1\le i \le N$.} with circulant
matrices, with respect to implementation via rectifier circuits of
depth 2 or unbounded depth. Precisely, the bound for the first
measure is $\Omega(N^2\log^{-10} N)$, and for the second it is
$\Omega(N^2\log^{-12} N)$.

In fact, the method has a potential for improvement of the above
bounds, which is of interest due to connection to the Zarankiewicz
problem (the problem is discussed in details e.g. in~\cite{ES}).
This potential is in application of a more accurate bound on the
cardinality of the sum of two sets in a Euclidean space following
from~\cite{gg,r}.

Below, we show the existence of $(k,l)$-free circulant $N\times N$
matrices of weight
$\Omega\left(\frac{k+l}{k^2l^2}N^{2-\frac{k+l+2}{kl}}\right)$. For
comparison, the classic Erd\"os---Spencer result~\cite{ES} states
just a slightly better bound
$\Omega_{k,l}\left(N^{2-\frac{k+l-2}{kl-1}}\right)$ in the class
of all $(k,l)$-free matrices.

Hence, for a system of Boolean sums with an appropriate circulant
matrix the following complexity bounds hold:

--- $\Omega(N^2\log^{-6} N)$ with respect to implementation via
circuits of functional elements\footnote{Further, we simply call
them circuits.} over the basis $\{\vee, \wedge\}$;

--- $\Omega(N^2\log^{-5}N)$ with respect to implementation via
circuits over the basis $\{\vee\}$, or via rectifier circuits;

--- $\Omega(N^2\log^{-4} N)$ with respect to implementation via depth-2
rectifier circuits.

The paper~\cite{norect} considers the ratio $\lambda(N) = \max_A
\frac {L_{\vee}(A)}{L_{\oplus}(A)}$, where $L_{\vee}(A)$ is the
circuit complexity of the Boolean sums' system with matrix $A$
over the basis $\{\vee\}$, $L_{\oplus}(A)$ is the circuit
complexity of the linear operator with matrix $A$ over the basis
$\{\oplus\}$, and the maximum is taken over all Boolean $N\times
N$ matrices. The result of the present paper leads to a bound
$\lambda(N) = \Omega\left( \frac N {(\log N)^6 \log\log N}
\right)$, which in a sense close to an upper bound $\lambda(N) =
O\left(\frac{N}{\log N}\right)$.

As another corollary, we obtain that the circuit complexity of the
Boolean convolution of order $N$ over the basis $\{\vee, \wedge\}$
is $\Omega(N^2\log^{-6} N)$. Specifically, this bound holds for
the number of disjunctors (that is, $\vee$-gates) in any monotone
circuit computing the convolution. Some recent papers
(e.g.~\cite{K,Bl}) mention the bound $\Omega(N^{3/2})$ as a
record, though a stronger bound follows from~\cite{G}
directly\footnote{The bound $\Omega(N^{3/2})$ corresponds to the
number of disjunctors in a monotone circuit (the survey~\cite{K}
is inaccurate at this point). However, the recent paper~\cite{Bl}
declares the same bound for the number of conjunctors
($\wedge$-gates; proof is omitted there).}. The obtained lower
bound is close to the trivial upper bound $O(N^2)$.

\section{Some properties of ``rectangles''}

Now, we present the main result following the proof strategy
from~\cite{G}. Let $k,l \in \mathbb N$, $2 \le k \le l$. Denote
$\mathbb Z_+ = \mathbb N \cup \{0\}$. We define {\it rectangle} as
an element of the set
$$R_{k,l} = \{ (x_1,\ldots,x_k,y_1,\ldots,y_l) \in \mathbb Z_+^{k+l} \mid \forall_{i \ne j} (x_i \ne x_j), \, \forall_{i\ne j} (y_i \ne y_j) \}.$$

Let $E = (a_1,\ldots,a_k,b_1,\ldots,b_l)$ be a rectangle. Let
$m(E) = |\{a_i+b_j \mid 1\le i \le k, \; 1\le j \le l \} |$ denote
the number of points in the rectangle $E$.

Consider the system $S(E)$ of linear equations
$$\{  x_r+y_s = x_u+y_v \mid a_r+b_s = a_u+b_v, \; 1\le r,\,u \le k, \; 1\le s,\,v \le l \}$$
over the field $\mathbb R$. The set of solutions constitutes a
linear subspace $T_E$ in $\mathbb R^{k+l}$. Let $n(E)$ be its
dimension. Let $C(E)$ denote the set of rectangles
$\{(x_1,\ldots,x_k,y_1,\ldots,y_l)\}$ satisfying $S(E)$ and
failing to satisfy any other equation $x_r+y_s = x_u+y_v$
(in~\cite{G}, $C(E)$ is called equivalence class).

We have to estimate the number of rectangles with bounded (by a
number $N$) coordinates and fixed number of points. An implicit
relation between the number of rectangles and the number of points
will be further established with the help of intermediate
parameter $n(E)$. First, we will count the number of rectangles
$E$ with a given value of $n(E)$. Next, we will derive relations
between $n(E)$ and $m(E)$.

To roughly estimate the number of rectangles with bounded
coordinates $0\le x_1,\ldots,y_l < N$ in $C(E)$ we use the
following lemma.

\begin{lemma}\label{1}
Let $N \in \mathbb N$. Then $|C(E) \cap \{0,\ldots,N-1\}^{k+l}|
\le N^{n(E)}$.
\end{lemma}

\proof The coordinates $x_1,\ldots,x_k,y_1,\ldots,y_l$ of a vector
from $T_E$ are defined by values of $n(E)$ free variables. There
are at most $N^{n(E)}$ ways to arrange such values, given that the
vector is from $C(E) \cap \{0,\ldots,N-1\}^{k+l}$. \qed

The second lemma estimates the number of classes with a given
value of $n(E)$. (We use notation $C_n^k$ for binomial
coefficients.)

\begin{lemma}\label{2}
Let $n \in \mathbb N$. Then $|\{ C(E) \mid n(E)=n \}| \le
C_{k^2l^2}^{k+l-n}$.
\end{lemma}
\proof The class $C(E)$ is uniquely defined by the system $S(E)$,
which in its turn is uniquely defined by a linearly independent
subsystem of $k+l-n$ equations. The number of such subsystems is
bounded from above by the number of ways to choose $k+l-n$
equations from $k^2l^2$ ones. \qed

Now, we manage to obtain relations between $n(E)$ and $m(E)$. This
piece of proof differs from~\cite{G}.

Let $\xi_i$ denote the unit vector in the space $\mathbb R^{k+l}$
with $i$-th coordinate being~1 and other coordinates being 0.

Let $n=n(E)$. For unification, let us introduce notation $x_{i+k}
= y_i$, $1 \le i \le l$. Set
$$\bar x = (\underbrace{1,\ldots,1}_k,\underbrace{0,\ldots,0}_l), \qquad \bar y =(\underbrace{0,\ldots,0}_k,\underbrace{1,\ldots,1}_l).$$
Notice that $\bar x, \bar y \in T_E$ (regardless of $E$). Let
$T'_E$ be the space of solutions of the system
$$ S'(E) = S(E) \cup \{x_1=y_1=0\}. $$
Then $\dim T'_E = n-2$ and $T_E = T'_E + \{\alpha \bar x + \beta
\bar y \mid \alpha,\beta \in \mathbb R\}$ ($A+B$ hereafter denotes
the element-wise sum (Minkowski sum) of sets $A$ and $B$). Write
$$ T'_E = \left\{ (x_1,\ldots,x_{k+l}) \left| \; x_i = \sum_{j=1}^{n-2} \alpha_{i,j} x_{i_j}, \; 1\le i \le k+l \right. \right\}, $$
where $x_{i_1},\ldots,x_{i_{n-2}}$ is a set of free variables of
the system $S'(E)$, and $\alpha_{i,j}$ are real constants. Then
setting $i_{n-1}=1$ and $i_n=k+1$ we conclude that
$x_{i_1},\ldots,x_{i_n}$ is a set of free variables of the system
$S(E)$, and
$$ T_E = \left\{ (x_1,\ldots,x_{k+l}) \left| \; x_i = \sum_{j=1}^n \alpha_{i,j} x_{i_j}, \; 1\le i \le k+l \right. \right\}, $$
where $\alpha_{i,n-1}=\alpha_{k+j,n}=1$, and
$\alpha_{i,n}=\alpha_{k+j,n-1}=0$ for $1\le i \le k$, $1 \le j \le
l$.

Consider a linear mapping $\psi_E$ from $\mathbb R^{k+l}$ to the
space $\mathbb R^{n-2}$ with Euclidean metrics and orthonormal
basis $\{e_1, \ldots, e_{n-2}\}$ defined by $\psi_E: \xi_i \to
\sum_{j=1}^{n-2} \alpha_{i,j} e_j$ for any $i$. In particular,
$\psi_E(\xi_1) = \psi_E(\xi_{k+1})=0$ ($0$ hereafter stands for
the zero vector of a space if it does not lead to a
misunderstanding).

Set $A_E=\psi_E (\{ \xi_1, \ldots, \xi_k \})$, $B_E=\psi_E (\{
\xi_{k+1}, \ldots, \xi_{k+l}\})$.

Recall that the dimension $\dim A$ of a set $A$ in a Euclidean
space is the minimum of dimensions of affine subspaces containing
$A$.

\begin{lemma}\label{3}
$|A_E+B_E|=m(E)$, $\dim (A_E+B_E)=n-2$.
\end{lemma}

\proof The first equality holds due to the following chain of
equivalent transformations:
$$ a_r + b_s = a_u+b_v  \eq $$
$$ \left(\; (x_1,\ldots,x_k,y_1,\ldots,y_l) \in T_E \Longrightarrow x_r + y_s = x_u+y_v \; \right) \eq  $$
$$ \left(\; (x_1,\ldots,x_k,y_1,\ldots,y_l) \in T_E \Longrightarrow
  \sum_{j=1}^n (\alpha_{r,j} + \alpha_{k+s,j}) x_{i_j} = \sum_{j=1}^n (\alpha_{u,j} + \alpha_{k+v,j}) x_{i_j} \; \right) $$
$$ \eq  \forall_{j, \; 1\le j \le n} (\alpha_{r,j} + \alpha_{k+s,j} = \alpha_{u,j} + \alpha_{k+v,j}) \eq $$
$$ \forall_{j, \; 1\le j \le n-2} (\alpha_{r,j} + \alpha_{k+s,j} = \alpha_{u,j} + \alpha_{k+v,j}) \eq $$
$$ \psi_E(\xi_r+\xi_{k+s}) = \sum_{j=1}^{n-2} (\alpha_{r,j} + \alpha_{k+s,j}) e_j =
 \sum_{j=1}^{n-2} (\alpha_{u,j} + \alpha_{k+v,j}) e_j = \psi_E(\xi_u+\xi_{k+v}). $$

The second equality is straightforward, since $0 \in A_E \cap B_E$
and $\{e_1, \ldots, e_{n-2}\} \subset A_E \cup B_E$. \qed

In the next section, we will estimate $m(E)$.

\section{The cardinality of the sum of two sets in a Euclidean space}

The following result is due to I. Ruzsa~\cite{r}.
\begin{theorem}[Ruzsa~\cite{r}]\label{R}
Let $A$ and $B$ be finite sets in the Euclidean space $\mathbb
R^n$ satisfying $|A|\le|B|$ and $\dim (A+B)=n$. Then
$$ |A+B| \ge n|A|+|B| - \frac{n(n+1)}2.  $$
\end{theorem}
Ruzsa also provided a more accurate bound
$$  |A+B| \ge |B| + \sum_{i=1}^{|A|-1} \min \{ n,\, |B|-i \}. $$

W.l.o.g. we can assume $0 \in A \cap B$ throughout this section.

Already, these bounds are sufficient to principally achieve
results announced in the introduction. However, the bounds are not
asymptotically tight for large $n$. On the contrary, the bound
$|A+B|\ge \lfloor n^2/4 \rfloor$ established in~\cite{G} is rough
for small $n$ (though its advantage is the simplicity of the
proof). The method~\cite{gg} allows to exhibit tight bounds.

Let $\{e_1, \ldots, e_n\}$ be an orthonormal basis of a Euclidean
space $\mathbb E^n$. Following~\cite{gg}, we define {\it long
simplex} as a set $F$ of the form
\begin{equation}\label{ls}
 \{ me_1 | m=0,\ldots, |F|-k \} \cup \{ e_{i_1}, \ldots, e_{i_{k-1}} \},
\end{equation}
with numbers $1, i_1, \ldots, i_{k-1}$ being pairwise different,
$k \ge 0$.

The next lemma is a reformulation of the Corollary 3.8~\cite{gg}.

\begin{lemma}\label{gg}
Under conditions of Theorem~$\ref{R}$ the minimum of $|A+B|$ is
either $|A||B|$ $($in this case $\dim A + \dim B = n)$, or it is
witnessed by a pair of long simplices.
\end{lemma}

The proof can be found in~\cite{gg}. It is crucial to observe that
the sets $A$ and $B$ delivering the minimum in the lemma satisfy
the definition~(\ref{ls}) with the same basis and, in particular,
with the same vector $e_1$.

Tight bounds (for any values of parameters) were not determined
in~\cite{gg}. Though, they can be easily derived from the lemma
above.
\begin{theorem}\label{dim}
Let $A, B \subset \mathbb R^n$, $K=|A|\le|B|=L$, and $\dim
(A+B)=n$. We have:

$(i)$ if $n=K+L-2$, then $ |A+B| = KL$;

$(ii)$ if $n \le L-K$, then $ |A+B| \ge L + n(K-1); $

$(iii)$ if $L-K \le n \le L$, then
$$ |A+B| \ge (n+1)K - \frac{(n-L+K)(n-L+K+1)}2;  $$

$(iv)$ if $L \le n \le K+L-3$, then
$$ |A+B| \ge KL - \frac{(K+L-n)(K+L-n-1)}2.  $$
\end{theorem}

\proof In the case $\dim A + \dim B = n$, the set $A+B$ has the
maximal possible cardinality $KL$, thus, $(i)$ follows. Therefore,
in the case $n < K+L-2$, we may assume that $\dim A + \dim B
> n$.

So, by Lemma~\ref{gg}, it suffices to consider sets $A$, $B$ being
long simplices~(\ref{ls}). Assume w.l.o.g.
$$ A = C_A \cup D \cup D_A, \qquad B = C_B \cup D \cup D_B, $$
where
$$ C_A =  \{ me_1 | m=0,\ldots, K-s-s_A-1 \},  $$
$$ C_B =  \{ me_1 | m=0,\ldots, L-s-s_B-1 \},  $$
$$ D = \{e_2,\ldots, e_{s+1}\}, \quad D_A = \{e_{s+2},\ldots, e_{s+s_A+1}\}, \quad D_B = \{e_{s+s_A+2},\ldots, e_n\},  $$
$s=|D|$, $s_A=|D_A|$, $s_B=|D_B|$, $s+s_A+s_B=n-1$. Hence,
\begin{multline*}
|A+B| = |C_A+C_B| + |(C_A \cup C_B)+D| + |C_A+D_B| + \\ |C_B+D_A|
+ |D+D| + |D + (D_A \cup D_B)| + |D_A+D_B|.
\end{multline*}
It can be verified directly that
$$ |C_A+C_B| = K+L-s-n, \quad |(C_A \cup C_B)+D| = s(\max\{L-s_B, \, K-s_A\}-s), $$
$$ |C_A+D_B| = (K-s-s_A)s_B, \quad |C_B+D_A| = (L-s-s_B)s_A,   $$
$$ |D+D| = \frac{s(s+1)}2, \quad |D + (D_A \cup D_B)| = s(s_A+s_B), \quad |D_A+D_B|=s_As_B.  $$
Summing all, we obtain
\begin{equation}\label{a+b}
|A+B| =
(s_A+1)K+(s_B+1)L+s\cdot\max\{L-s_B,\,K-s_A\}-n-\frac{s(s+1)}2-s_As_B.
\end{equation}
Thus, the problem reduced to finding the minimum of the
expression~(\ref{a+b}). Let $s^*$, $s_A^*$, $s_B^*$ denote the
values of parameters $s$, $s_A$, $s_B$ delivering this minimum.
Let us list restrictions on the parameters:
$$ s+s_A+s_B=n-1, \qquad s+s_A \le K-1, \qquad s+s_B \le L-1. \eqno(*)$$

Consider $(ii)$. Suppose $n \le L-K$. Then
$$ L-s_B \ge K + (n-s_B) \ge K \ge K-s_A.$$
Thus, minimization of~(\ref{a+b}) (with eliminated constant terms)
is equivalent to maximization of the expression
\begin{equation}\label{s_B}
s_B(n+L-K-1-s_B) + \frac{s(s+1)}2.
\end{equation}
For a fixed $s$ the value of~(\ref{s_B}) grows when $s_A$
decreases (and $s_B$ increases accordingly), since $2s_B <
n+L-K-1$ and due to the fact that the function $x(a-x)$
monotonically grows in the interval $[0,\,a/2]$. Yet, the
conditions $(*)$ are not violated. Hence, $s_A^*=0$.

Set $s_B=n-1-s$. Then, after elimination of constant terms the
expression~(\ref{s_B}) reduces to
$$ -\frac{s(s+1)}2 -((L-K)-n)s. $$
Consequently, $s^*=0$. By the assignment $s_B=n-1$ and $s=s_A=0$
in~(\ref{a+b}), we derive the inequality $(ii)$.

Let us prove $(iii)$. Assume $L-K \le n \le L$. Consider two
cases.

Case A. Suppose $L-s_B \ge K-s_A$. As above, the problem reduces
to maximization of~(\ref{s_B}). Note that for a fixed $s_B$ the
value of~(\ref{s_B}) grows with decreasing of $s_A$ (and
corresponding increasing of $s$), and the conditions $(*)$ are not
violated. Therefore, either $s^*_B \le L-K$ and $s^*_A=0$, or
$s^*_B=L-K+s_A^*$.

In the former subcase, assign $s=n-1-s_B$. Then, after elimination
of constant terms the expression~(\ref{s_B}) reduces to
$$  s_B(2(L-K)-1-s_B), $$
hence, $s_B^* \in \{L-K-1,\,L-K\}$.

In the latter subcase, assign $s_B=L-K+s_A$ and $s=n-1-L+K-2s_A$.
Then, the expression~(\ref{s_B}) has the form
$$ s_A(s_A+L-K-n). $$
The second factor is $s_B-n$, and so it is negative. Consequently,
$s_A^*=0$, and $s^*_B = L-K$ follows as well, as in the previous
subcase.

Case B. Suppose $L-s_B \le K-s_A$. Then,
$$ s+s_A = n-1-s_B \le L-1-s_B \le K-1-s_A \le K-1. $$
So, only the first of conditions $(*)$ is essential. Here,
minimization of~(\ref{a+b}) is equivalent to maximization of the
expression
\begin{equation}\label{s_A}
s_A(n-L+K-1-s_A) + \frac{s(s+1)}2.
\end{equation}
For a fixed $s_A$ the value of~(\ref{s_A}) grows, when $s$
increases and $s_B$ accordingly decreases, thus,
$s_B^*=L-K+s_A^*$. That is the very situation already discussed in
the second subcase of the case A.

Via assignment $s_A=0$, $s_B=L-K$, $s=n-1-L+K$ in~(\ref{a+b}), we
obtain the inequality $(iii)$ (the assignment is in a sense
correct also in the case $L-K=n$).

Now, turn to $(iv)$. Assume $L \le n \le K+L-3$. Again, consider
two cases.

Case A. Suppose $L-s_B \ge K-s_A$. In this case, the latter of
conditions $(*)$ follows from the second:
$$ s+s_B \le s+s_A + L - K \le L-1. $$
Again, the problem is to maximize the expression~(\ref{s_B}).
Observe that for a fixed $s_B$ the value of~(\ref{s_B}) grows when
$s_A$ decreases (and $s$ correspondingly increases), and
conditions $(*)$ are not violated. Hence, $s^*_A = s^*_B-L+K$ (it
is the minimal possible value of $s_A$ for a fixed $s_B$).

Under the assignment $s=n+L-K-1-2s_B$ and elimination of constant
terms, the expression~(\ref{s_B}) reduces to
$$ s_B(s_B-n-L+K). $$
Since $2s_B < (L-K+s_A)+(n-s_A) = n+L-K$, the second factor is
negative and greater than the first factor by absolute value.
Consequently, the maximum is achieved on the minimal possible
value of $s_B$ under the conditions $(*)$. Hence, we deduce that
$s_B^*=n-K$.

Case B. Suppose $L-s_B \le K-s_A$. In this case, the second
condition in $(*)$ is inessential:
$$ s+s_A \le s+s_B - L + K \le K-1. $$
We have to maximize~(\ref{s_A}). Observe that it grows when $s_A$
is fixed, $s$ increases and $s_B$ decreases, and conditions $(*)$
are fulfilled. Thus, $s_B^*=L-K+s_A^*$. So, we are under the
conditions of the already investigated case A.

Under assignment $s_A=n-L$, $s_B=n-K$, $s=L+K-1-n$ in~(\ref{a+b}),
we exhibit the inequality $(iv)$. \qed

As follows from the proof, the bounds of the theorem are
achievable.

Under the conditions of Theorem~\ref{dim}, define the function
\begin{equation}\label{rho}
\rho(K,L) = \max_{1\le n \le K+L-2} \frac{n+2}{|A+B|}.
\end{equation}

\begin{lemma}\label{prec}
$\rho(2,L) = \frac{L+2}{2L}$. If $K\ge 3$, then
$$ \rho(K,L) = \max \left\{ \frac{K+L}{KL},\;  \frac{K+L-1}{KL-3},\;  \frac{2(L+2)}{K(2L-K+1)} \right\} < \frac{K+L+2}{KL}. $$
In particular, $\rho(K,K) = \frac{2(K+2)}{K(K+1)}$.
\end{lemma}

\proof Define additionally $\rho(K,L,n) =
\frac{n+2}{\min_{A,B}|A+B|}$. By the definition, $\rho(K,L) =
\max_n \rho(K,L,n)$.

First, we need to verify that the function $\rho(K,L,n)$ achieves
its maximum at the endpoints of intervals defined in pp.
$(ii)$--$(iv)$ of Theorem~\ref{dim}.

In the case $1 \le n \le L-K$, the function
$$\rho^{-1}(K,L,n) = \frac{L+n(K-1)}{n+2} = K-1 + \frac{L-2K+2}{n+2} $$
is evidently monotone (hereafter, we consider $\rho(K,L,n)$ as a
function of variable~$n$).

In the case $L-K \le n \le L$, denote $n'=n-(L-K)$. Then
$$\rho^{-1}(K,L,n) = K - \frac{n'(n'+1)+2K}{2(n'+L-K+2)}.$$
The subtrahend function is convex downward for $n'\ge0$, since it
has the form $c\frac{n'(n'+1)+a}{n'+1+b}$ with $a,b,c\ge0$.
Therefore, with respect to the interval $[0,\,K]$ it takes its
maximal value in the endpoints (it holds for $K\ge 3$; for $K=2$
the argument of the maximum lies in the interval $[0,\,1]$).
Consequently, there takes its maximum the function $\rho(K,L,n)$.

In the case $L \le n \le K+L-3$, denote $n'=n-L$. Then
\begin{multline*}
\rho^{-1}(K,L,n) = K - \frac{2(n'+2)K+(K-n')(K-n'-1)}{2(n'+L+2)}=
\\ = K - \frac{n'(n'+1)+K(K+3)}{2(n'+L+2)}.
\end{multline*}
We treat this case the same way as the previous one.

Thus, for $K\ge3$ we have
$$ \arg \max_{1\le n \le K+L-2} \rho(K,L,n) \in \{ 1,\, L-K,\, L,\, K+L-3,\, K+L-2 \}, $$
$$ \arg \max_{1\le n \le K+L-2} \rho(2,L,n) \in \{ 1,\, L-2,\, L-1,\, L \}. $$

Let us check that $\rho(K,L,1) \le \rho(K,L,K+L-2)$. Indeed,
$$ \rho(K,L,1) = \frac{3}{K+L-1} \le \frac{4}{K+L} \le \frac1K + \frac1L = \frac{K+L}{KL} = \rho(K,L,K+L-2), $$
due to the well-known inequality $\frac{a^2}{b}+\frac{c^2}{d} \ge
\frac{(a+c)^2}{b+d}$, where $b,d>0$.

Notice further that
$$ \rho(K,L,L-K) = \frac1K\left(1+ \frac1{L-(K-1)}\right) \le \frac1K\left(1+ \frac{K}{L}\right) = \rho(K,L,K+L-2). $$

Yet,
$$ \rho(2,L,L-1) = \frac{L+1}{2L-1} \le \frac{L+2}{2L} = \rho(2,L,L).  $$

Therefore, it is proved that $\rho(2,L) = \rho(2,L,L) =
\frac{L+2}{2L}$ and
\begin{multline*}
 \rho(K,L) = \max \left\{ \rho(K,L,K+L-2),\, \rho(K,L,K+L-3),\, \rho(K,L,L) \right\} = \\ \max \left\{ \frac{K+L}{KL},\;
\frac{K+L-1}{KL-3},\;  \frac{2(L+2)}{K(2L-K+1)} \right\}.
\end{multline*}

Applying the simple estimation
$$ \frac{2(L+2)}{K(2L-K+1)} = \frac{L+(K+3)\frac{L}{2L-K+1}}{KL} \le \frac{L+(K+3)\frac{K}{K+1}}{KL} < \frac{K+L+2}{KL}, $$
the inequality $\rho(K,L) < \frac{K+L+2}{KL}$ can be easily
checked. The last statement of the lemma concerning $\rho(K,K)$ is
easy to verify. \qed

\section{Weight of thin circulant matrices}

A circulant matrix is entirely defined by its one row, say, the
first row. Let $c_j=c_{0,j}$, $0\le j \le N-1$, denote the entries
of the row, where $N$ is the size of the matrix. For convenience,
assume that the other entries satisfy $c_{i,j} = c_{(i+j) \bmod
N}$ (that is, 1-uniform diagonals of the matrix are parallel to
the secondary diagonal).

Then, the condition that a matrix $(c_{i,j})$ contains an all-ones
submatrix constituted by rows with numbers $a_1,\ldots,a_k$ and by
columns with numbers $b_1,\ldots,b_l$ can be written as
$$ c_{(a_i+b_j) \bmod N} = 1, \qquad 1\le i\le k, \; 1\le j\le l. $$

Let $\gamma_0,\ldots,\gamma_{N-1}$ be independent random variables
taking value 1 with probability $p$ and value 0 with probability
$1-p$. Denote $\gamma = \sum \gamma_i$.

Hereafter, we denote by ${\bf P}(Q)$ the probability of the event
$Q$. Let ${\bf M} \xi$ and ${\bf D} \xi$ denote the expectation
and the variance of a random variable $\xi$, respectively.

\begin{lemma}\label{P}
${\bf P} \left(\gamma \ge pN-2\sqrt{pN}\right) \ge 3/4. $
\end{lemma}
\proof The required inequality follows from the Chebyshev's
inequality
$$ {\bf P} \left( |\gamma - {\bf M}\gamma| > \varepsilon \right) < \frac{{\bf D}\gamma}{\varepsilon^2} $$
by setting ${\bf M}\gamma = pN$, ${\bf D}\gamma = p(1-p)N$ è
$\varepsilon = 2\sqrt{pN}$. \qed

Set formally $\gamma_i=0$, when $i \ge N$. Let
$Q(E,\gamma_0,\ldots,\gamma_{N-1})$ with $E =
(a_1,\ldots,a_k,b_1,\ldots,b_l) \in R_{k,l}\cap
\{0,\ldots,N-1\}^{k+l}$ denote the event
$$ \forall_{i,\,j%: \; 1\le i\le k,\, 1\le j\le l
                 } (\gamma_{a_i+b_j}=1). $$
Substantially, it implies that a random circulant $2N \times 2N$
matrix $\Gamma$ with the first row
$(\gamma_0,\ldots,\gamma_{N-1},0,\ldots,0)$ contains an all-ones
$k\times l$ submatrix in the intersection of rows $a_1,\ldots,a_k$
and columns $b_1,\ldots,b_l$.

Observe that any all-ones submatrix of a matrix $\Gamma$ can be
translated to an all-ones submatrix entirely contained in the
upper left $N \times N$ submatrix (that is, constituted by rows
and columns numbered from 0 to $N-1$) of $\Gamma$ by a cyclic
shift (of numbers of rows and columns). Generation of all-ones
submatrices by cyclic shifts is illustrated on the picture below;
submatrices $C_i$ are shown as rectangles, the submatrix $C_0$ is
a desired one.

\begin{figure}[htb]
\begin{picture}(200,200)(-90,0)

\thicklines

\multiput(0,0)(0,200){2}{\line(1,0){200}}
\multiput(0,0)(200,0){2}{\line(0,1){200}}
\multiput(100,0)(-100,100){2}{\line(1,1){100}}
\put(0,0){\line(1,1){200}}

\thinlines

\multiput(110,45)(0,40){2}{\line(1,0){30}}
\multiput(110,45)(30,0){2}{\line(0,1){40}}

\multiput(180,115)(0,40){2}{\line(1,0){20}}
\multiput(0,115)(0,40){2}{\line(1,0){10}}
\multiput(10,115)(170,0){2}{\line(0,1){40}}

\multiput(60,35)(0,160){2}{\line(1,0){30}}
\multiput(60,0)(30,0){2}{\line(0,1){35}}
\multiput(60,195)(30,0){2}{\line(0,1){5}}

\multiput(15,150)(0,40){2}{\line(1,0){30}}
\multiput(15,150)(30,0){2}{\line(0,1){40}}

\put(95,0){\line(1,1){105}} \put(65,0){\line(1,1){135}}
\put(55,0){\line(1,1){145}} \put(25,0){\line(1,1){175}}

\put(0,105){\line(1,1){95}} \put(0,135){\line(1,1){65}}
\put(0,145){\line(1,1){55}} \put(0,175){\line(1,1){25}}

\put(75,125){$0$} \put(170,25){$0$}

\put(18,179){$C_0$} \put(63,24){$C_1$} \put(113,74){$C_2$}
\put(183,144){$C_3$}

\end{picture}
%\caption{}
\end{figure}

Therefore, the matrix $\Gamma$ is $(k,l)$-free iff its left upper
$N\times N$ submatrix is.

\begin{theorem}\label{main}
There exists a $(k,l)$-free circulant $N\times N$ matrix of
weight $\Omega\left(\frac{k+l}{k^2l^2}N^{2-\rho(k,l)}\right)$.
\end{theorem}

\proof It follows directly from the definition that the
probability of the event $Q(E,\gamma_0,\ldots,\gamma_{N-1})$ is at
most $p^{m(E)}$. Then
\begin{multline*}
{\bf P} \left( \exists_{E \in R_{k,l}}
(Q(E,\gamma_0,\ldots,\gamma_{N-1})) \, \right) \le \\
\sum_{E \in R_{k,l} \cap \{0,\ldots,N-1\}^{k+l}} {\bf P}
\left(Q(E,\gamma_0,\ldots,\gamma_{N-1}) \, \right) = \\
\sum_{n=3}^{k+l} \sum_{ \begin{array}{c} \scriptstyle E \in
R_{k,l} \cap \{0,\ldots,N-1\}^{k+l}, \\ \scriptstyle n(E)=n
\end{array}} {\bf P} \left(Q(E,\gamma_0,\ldots,\gamma_{N-1}) \,
\right) \le \\
\sum_{n=3}^{k+l} \sum_{ \begin{array}{c} \scriptstyle E \in
R_{k,l} \cap \{0,\ldots,N-1\}^{k+l}, \\ \scriptstyle n(E)=n
\end{array}} p^{m(E)} \le
\sum_{n=3}^{k+l} \sum_{\begin{array}{c} \scriptstyle C(E) \subset R_{k,l}, \\
\scriptstyle n(E)=n
\end{array}} N^np^{m(E)}
\le \\
\sum_{n=3}^{k+l} \sum_{\begin{array}{c} \scriptstyle C(E) \subset R_{k,l}, \\
\scriptstyle n(E)=n
\end{array}}
\left(pN^{\rho(k,l)}\right)^{m(E)}.
\end{multline*}
Here, the second from the last inequality follows from
Lemma~\ref{1}, and the last one is justified by Lemma~\ref{3} and
the definition~(\ref{rho}).

Set $p=\left(\frac{k+l}{ek^2l^2}\right)N^{-\rho(k,l)}$, and
continue exploiting the inequality of Lemma~\ref{2}:
\begin{multline*}
\sum_{n=3}^{k+l} \sum_{\begin{array}{c} \scriptstyle C(E) \subset R_{k,l}, \\
\scriptstyle n(E)=n
\end{array}}
\left(pN^{\rho(k,l)}\right)^{m(E)} \le \sum_{n=3}^{k+l} \sum_{\begin{array}{c} \scriptstyle C(E) \subset R_{k,l}, \\
\scriptstyle n(E)=n
\end{array}} \left(\frac{k+l}{ek^2l^2}\right)^{m(E)} \le \\
\sum_{n=3}^{k+l} C_{k^2l^2}^{k+l-n}
\left(\frac{k+l}{ek^2l^2}\right)^{k+l-1} \le \sum_{n=3}^{k+l}
\left(\frac{ek^2l^2}{k+l-n}\right)^{k+l-n}
\left(\frac{k+l}{ek^2l^2}\right)^{k+l-1} = \\
\sum_{n=3}^{k+l} \left(\frac{k+l}{ek^2l^2}\right)^{n-1} \left( 1 +
\frac{n}{k+l-n}\right)^{k+l-n} \le  \sum_{n=3}^{k+l}
\left(\frac{k+l}{ek^2l^2}\right)^{n-1} e^n = \\
e \sum_{n=3}^{k+l} \left(\frac{k+l}{k^2l^2}\right)^{n-1} \le
 \frac{e(k+l)}{k^2l^2} \le e/4.
\end{multline*}
Here, we use well-known inequalities $C_n^m \le \left( \frac{en}m
\right)^m$ and $(1+1/x)^x < e$ for $x>0$, and assume $x^x
\mid_{x=0} = 1$ (this quantity appears in the form
$(k+l-n)^{k+l-n} \mid_{n=k+l}$).

Hence, as follows form the note before the theorem, a random
circulant ($2N\times 2N$) matrix $\Gamma$ is $(k,l)$-free with
probability at least $(4-e)/4$. In the sight of Lemma~\ref{P}, we
can conclude that this random matrix is $(k,l)$-free and also has
weight $2N\gamma \ge 2N(pN - 2\sqrt{pN})=\Omega(pN^2)$ with
positive probability. \qed

\section{Corollaries}

Theorem~\ref{main} and Lemma~\ref{prec} lead to
\begin{corol}
There exists a $(k,l)$-free $N\times N$ circulant matrix of weight
$\Omega\left(\frac{k+l}{k^2l^2}N^{2-\frac{k+l+2}{kl}}\right)$.
\end{corol}

In the case $k=l=\Theta(\log N)$, the weight of a circulant matrix
provided by the corollary is $\Omega\left(N^2\log^{-3} N\right)$.
This fact together with complexity bounds for Boolean sums'
systems with $(k,l)$-free matrices~\cite{m} (see also~\cite{G,W})
yields

\begin{corol}\label{c2}
There exists a circulant $N \times N$ matrix such that for the
complexity of the corresponding system of Boolean sums the
following bounds hold: $\Omega\left(N^2\log^{-4} N\right)$ with
respect to implementation via depth-$2$ rectifier circuits,
$\Omega\left(N^2\log^{-5}N\right)$~--- for circuits over the basis
$\{\vee\}$ or unbounded-depth rectifier circuits,
$\Omega\left(N^2\log^{-6} N\right)$~--- for the number of
disjunctors in a circuit over the basis $\{\vee, \wedge\}$.
\end{corol}

For the same choice of the parameters, the function $\lambda(N)$
defined in the introduction can be bounded as follows
(taking~\cite{norect} into account).
\begin{corol}
$\lambda(N) = \Omega\left( \frac N {(\log N)^6 \log\log N}
\right)$.
\end{corol}

{\it Boolean convolution of order} $N$ is the function
$$U_N(x_0,\ldots,x_{N-1},y_0,\ldots,y_{N-1}) = (u_0,\ldots,u_{2N-2}), \quad u_k = \bigvee_{i+j=k} x_iy_j. $$
{\it Cyclic Boolean convolution of order} $N$ is defined as
$$Z_N(x_0,\ldots,x_{N-1},y_0,\ldots,y_{N-1}) = (z_0,\ldots,z_{N-1}), \quad z_k = \bigvee_{i+j \equiv k \bmod N} x_iy_j. $$
Let $V(f)$ be the minimal number of disjunctors in a circuit over
the basis $\{\vee,\,\wedge\}$ that implements a function $f$.
Then, the following relations are straight from the definition of
convolutions:
$$ V(Z_N) \le V(U_N)+N-1, \qquad  V(U_N) \le V(Z_{2N-1}). $$

A cyclic Boolean convolution (up to a permutation of its
components) can be viewed as a system of Boolean sums of arguments
$x_0,\ldots,x_{N-1}$ with a variable circulant matrix defined by
the row $y_{N-1},\ldots,y_0$. Since the complexity of a circuit
(here, in the sense of the complexity measure $V(f)$) does not
increases after a replacement of some inputs by constants, we can
conclude that the complexity of the cyclic convolution of order
$N$ is at least the complexity of a system of Boolean sums with an
arbitrary circulant $N\times N$ matrix. So, by Corollary~\ref{c2},
we obtain
\begin{corol}
$V(U_N),\, V(Z_N) = \Omega\left(N^2\log^{-6} N\right).$
\end{corol}

Grinchuk Mikhail Ivanovich, e-mail: {\sf grinchuk@nw.math.msu.su}

Sergeev Igor Sergeevich, e-mail: {\sf isserg@gmail.com}

\section*{Notes (2017)}

By now, $\lambda(N)$ is proven to be $\Omega(N/\log^2 N)$. There
are several ways to show it, see e.g. [Jukna S., Sergeev I.
Complexity of linear boolean operators. Foundations and Trends in
Theoretical Computer Science. 2013. V. 9(1). 1--123] and
references there.

An explicit circulant matrix $A$ achieving
$L_{\vee}(A)/L_{\oplus}(A) = N^{1-o(1)}$ was constructed in
[Gashkov S. B., Sergeev I. S. A method for deriving lower bounds
for the complexity of monotone arithmetic circuits computing real
polynomials. Sbornik: Mathematics. 2012. V. 203(10), 1411--1447]
with the use of a combinatorial result by J. K\'ollar, L. R\'onyai
and T. Szab\'o.

\end{document}